\documentclass[referee]{aa} 
\usepackage{graphicx}
\usepackage{txfonts}
\usepackage{natbib}
 \usepackage[modulo,switch]{lineno}
\usepackage{multirow}

\bibpunct{(}{)}{;}{a}{}{,} 

\usepackage{graphicx}
 \usepackage{float}
\usepackage{txfonts}

\begin{document}

   \title{Photometric variability of the Be star CoRoT\thanks{The CoRoT  space mission was developed and is operated by the French space agency CNES, with participation of ESA's RSSD and Science Programmes, Austria, Belgium, Brazil, Germany, and Spain}-ID 102761769}

   \titlerunning{Photometric variability of the Be star CoRoT-ID 102761769}

   \authorrunning{Emilio et al.}


   \author{M. Emilio\inst{1},
          L. Andrade\inst{2},
          E. Janot-Pacheco\inst{2},
          A. Baglin\inst{3},
          J. Guti\'errez-Soto\inst{4},
          J.C. Su\'arez\inst{5}
          B. de Batz\inst{6},
          P. Diago\inst{4,5},
          J. Fabregat\inst{4},
          M. Floquet\inst{6},
          Y. Fr\'emat\inst{7},
          A.L. Huat\inst{6},
          A.M. Hubert\inst{6},
          F. Espinosa Lara\inst{6},
          B. Leroy \inst{3},
          C. Martayan\inst{6,8},
          C. Neiner \inst{6},
          T. Semaan\inst{6}
          \and
          J. Suso\inst{4}
          }

   \institute{Observat\'orio Astron\^omico/DEGEO, Universidade Estadual de Ponta Grossa,
              Av. Carlos Cavalcanti, 4748 Ponta Grossa, PR, Brazil - 84030-900, email:memilio@uepg.br
         \and
             Universidade de S\~ao Paulo/IAG-USP, rua do Mat\~ao, 1226 - Cidade Universit\'aria
S\~ao Paulo, SP, Brazil - 05508-900
         \and
             Observatoire de Meudon  LESIA, Observatoire de Paris, CNRS, Universit\'e Paris Diderot; 5 place Jules Janssen, 92190 Meudon, France
         \and
          Observatori Astron\`omic de la Universitat de Val\`encia, Edifici Instituts d'Investigaci\'o, Pol\'igon La Coma, 46980 Paterna, Val\`encia, Spain.
          \and
             Instituto de Astrof\'isica de Andaluc\'ia (CSIC), CP 3004, Granada, Spain
         \and
             Observatoire de Meudon  GEPI, Observatoire de Paris, CNRS, Universit\'e Paris Diderot; 5 place Jules Janssen, 92190 Meudon, France
         \and
             Royal Observatory of Belgium, 3 avenue circulaire, B1180 Brussels, Belgium
         \and
	     European Organization for Astronomical Research in the Southern Hemisphere, Alonso de Cordova 3107, Vitacura, Santiago de Chile, Chile
             }

   \date{Received January 16, 2010; accepted June 06, 2010}


  \abstract
   {Classical Be stars are rapid rotators of spectral type late O to early A and luminosity class V-III, wich exhibit Balmer
   emission lines and often a near infrared excess originating in an equatorially concentrated circumstellar envelope, both
   produced by sporadic mass ejection episodes. The causes of the abnormal mass loss (the so-called Be phenomenon)
   are as yet unknown. }
   {For the first time, we can now study in detail Be
   stars outside the Earth's atmosphere with sufficient temporal resolution. We investigate the variability of the Be Star CoRoT-ID 102761769
   observed with the CoRoT satellite in the exoplanet field during the initial run.}
   {One low-resolution spectrum of the star was obtained with the INT telescope at the Observatorio del Roque de los
    Muchachos. A time series analysis was performed using both cleanest and singular spectrum analysis algorithms to the
    CoRoT light curve. To identify the pulsation modes of the observed frequencies, we computed a set of
    models representative of CoRoT-ID 102761769 by varying its main physical parameters inside the uncertainties discussed.}
   {We found two close frequencies related to the star. They are 2.465 $\rm c\,d^{-1}$ (28.5 $\mathrm{\mu Hz}$) and
   2.441 $\rm c\,d^{-1}$ (28.2 $\mathrm{\mu Hz}$). The precision to which those frequencies were found is 0.018 $\rm c\,d^{-1}$
   (0.2 $\mathrm{\mu Hz}$). The projected stellar rotation was estimated to be 120 $\rm km\,s^{-1}$ from the Fourier transform of spectral lines. If CoRoT-ID 102761769 is a typical Galactic Be star it rotates near the critical velocity. The critical rotation frequency of a typical B5-6 star is about 3.5 $\rm c\,d^{-1}$(40.5 $\mathrm{\mu Hz}$),
    which implies that the above frequencies are really caused by stellar pulsations rather than star's rotation. }
   {}


   \maketitle

%

\section{Introduction}

Classical Be stars are rapid rotators of spectral type late O to early A and luminosity class V-III, wich exhibit Balmer emission lines and often a near infrared excess originating in an equatorially concentrated circumstellar envelope, both produced by sporadic mass ejection episodes. The causes of the abnormal mass loss (the so-called Be phenomenon) are as yet unknown. In spite of their high V $\sin i$ \citep{fre05}, rapid rotation alone cannot explain the ejection episodes as most Be stars do not rotate at their critical rotation rates \cite[see][for a comprehensive review of Be stars]{por03}.  \\
High-resolution, high signal-to-noise spectroscopic observations have been analyzed to demonstrate short-term variations are rather common among early-type Be stars \citep{baa82, baa84, por88, flo92, nei02}. The observed line profile variability (LPV) is characterized by moving bumps traveling from blue to red across the line profile on timescales ranging from minutes to a few days. The phenomenon has also been observed in O stars and $\delta$ Sct variables \citep[e.g.][]{ken92}, among others. Non-radial pulsations (NRP) have been proposed as an explanation of the LPV observed in hot stars \citep[e.g.][]{smi77, baa82, vog83}. NRP produce LPV thanks to the combination of the Doppler displacement of stellar surface elements with their associated temperature variations due to the compression/expansion caused by the passage of waves through the photosphere. NRP could be the additional mechanism required for a rapidly rotating B star to become a Be star, that is to trigger the Be phenomenon by means of mass ejection.  Indeed, \citet{riv98} found multiperiodicity in $\mu$ Cen and showed that a correlation exists between mass ejection episodes and the beating pattern of the multiperiodicity. The finding of new cases of Be stars for which beating periods of multiperiodic NRPs coincide with matter ejections would help us to confirm this model. The periodic variability of the star has been reproduced in detail by NRP modeling \citep{riv01}. Short-periodic LPV of other Be stars have also been modeled using NRP. Observations with the MOST satellite showed that multiperiodicity due to NRP is a rather common phenomenon among Be stars \citep[e.g.][]{wal05,dzi07}. Nevertheless, to measure frequencies with great accuracy in stars requires both high photometric precision and high time-frequency resolution. The CoRoT (Convection, Rotation and planetary Transits) satellite  opens for the first time in history the possibility of fulfilling this goal. CoRoT \citep{boi04} is an experiment of astronomy dedicated to seismology and the detection of extrasolar planet transits. It was launched on December 2006 in an inertial polar orbit at an altitude of 897 $\mathrm{km}$. The instrument is fed by a $\phi=27 \textrm{cm}$ telescope.  Its scientific program is three-fold consisting of: (1) The seismology core program (SISMO), which concerns the seismic study of $\sim$10 bright $(6<V<9)$ well-selected pulsating stars per observed field. Those measurements are sampled with a cadence of 32 seconds, which allows very accurate frequency analysis. (2) The exoplanet core program (EXO), which concerns the search for exoplanets around $\sim$12000 faint $(11.5<V<16)$ stars per field. The sampling of the EXO data has a cadence of 512 seconds. (3) The additional program (AP), which concerns the study of a few hundreds faint $(11.5<V<16)$ variable stars per EXO field. Within the core exoplanet data, there are hot stars among which there will certainly be new pulsators, enriching the statistics of the NRP phenomenon. In this work we report on the analysis of the star CoRoT-ID 102761769. Because of a certain instability of the attitude control system (satellite jitter), the analysis of some of CoRoT stars requires a careful treatment. In the next section, we present the light curve as coming from the reduction pipeline and explain the measured instrumental noise. In Sect. 3, we discuss the time analysis of the light curve and how the information was extracted. In Sect. 4, we describe the spectroscopy ground-based observations of the star and an estimate of the rotation period. In Sect. 5, we discuss the frequencies found and compare them with the critical rotation velocity. In Sect. 6, the conclusions of this work are presented.

\section{CoRoT photometric observations}

The data were taken in the exo field during the initial run (IRa01) during 54.6 days with an initial cadence of 512 seconds. An automatic procedure (alarm mode) triggers a higher cadence in the case of suspect detection of a planetary transit. After 3.13 days, CoRoT began to record the data of the ID 102761769 star with a cadence of 32 seconds (Fig. 1-top). The coordinates (J2000.0) of the star are RA=$06^{h}$ $45^{m}$ $02^{s}.4$ and DEC=$-01^{\circ}$ $50'$ $08".03$ and $V$= 13.288 mag. The star is also cited in catalogues mentioning as emission-line object \citep{koh99, rob89}.  The light curve contains 129\,305 photometric observations. Looking for oscillations, we first applied the cleanest algorithm \citep{fos95} to the data. Figure 2 shows the first cleanest power spectrum. We identified several frequencies, the most easily distinguishable ones being related to the satellite's orbit and its harmonics. The close distance from Earth causes perturbations on the satellite leading to satellite jitter \citep{auv09}. The main sources are eclipses, the variation in the Earth's gravity and magnetic field, South Atlantic anomaly, the Sun and Earth emissivity, Earth's albedo, and objects in low Earth orbit (LEO). Figure 3 shows a phase diagram of the raw data with the 6184 seconds CoRoT satellite orbit period. CoRoT pipeline have a flag for bad points. Some of the points were corrected by the pipeline, while others were eliminated, which is why we have missing points. The gaps in Fig. 3 were caused by points eliminated from the automatic pipeline reduction mainly by satellite jitter. The cluster of points in Fig. 3 are the remaining points that the pipeline could not correct. We eliminated 480 points greater than 3 $\sigma$ from the average (0.4\% of the total points).  We also eliminated the first 355 observations (0.3\% of the total points) taken with a larger cadence and filled the gaps with the average value (Fig. 1 - middle). The lower frequencies shown in the power spectrum in Fig. 2 are due to hot pixels \citep{auv09}. There is a prism mounted in front of the CoRoT exoplanet channel camera detectors, which allows CoRoT to gather 3-colour light curves. We analyzed the total sum of the light curves (the so-called white light curve) to highest quality relation in terms of signal/noise. Each color light curve falls on different pixels, what allows us to check for cosmic rays. The bump around date 2600 (Fig. 1 - middle) is only found in the red channel, and is indicative of cosmic-ray source noise. The lower frequencies were eliminated by a polynomial fit to the data and the missing points were interpolated (17715 in total). An average of each 100 points was finally evaluated  by reusing each point 100 times and then selecting 1 point every 100 points for additional analysis (Fig. 1 - bottom). The total number of points in Fig. 1 (bottom) is 1466. The regulary spaced intervals of data series were essential to the time analysis described in the next section.

\begin{figure*}[htp]
\centering
\includegraphics[width=0.6\textwidth,angle=0]{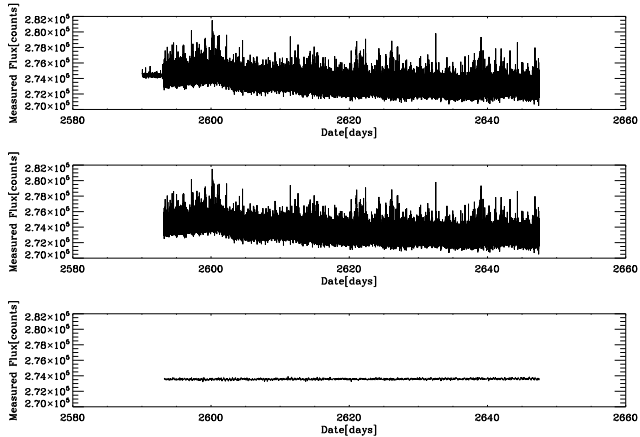}
\caption{Top:CoRoT-ID 102761769 star light curve. Middle: Data after outlier removal at three $\sigma$ level. Bottom: Moving average every 100 points.}
\label{fig01}
\end{figure*}

\section{Time analysis}

We applied the singular spectrum analysis (SSA) \citep{ghi02} to the data in Fig. 1-bottom. SSA is designed to extract information from noisy and evenly spaced time series. The method provides insight into the unknown or only partially known dynamics of the underlying system that generated the series. The procedure takes an univariate record and produces a multivariate set of observations by using lagged copies of a single time record, we are able to define the coordinates of the phase space that approximate the dynamics of the system. SSA has been applied extensively to the study of climate variability as well as other areas in physical and life sciences. To create the singular spectrum we have to sort the eigenvalues of the trajectory matrix. Each eigenvalue is associated with an oscillation in the time series. The full line in Fig. 4 (top) shows the singular spectrum applied to the Be star CoRoT-ID 102761769. This plot is the strength of the sorted eigenvalues. We performed a 100 runs of Monte Carlo SSA shuffling the data \citep{all96}. The dashed line in Fig. 4 represents the $3 \sigma$ confidence limit for this analysis. This means that the first four eigenvalues, defined by our Monte Carlo analysis,  almost entirely encompass the signal of our time series. We reconstructed the time series (filtered series) with this limit ($3^{rd}$ row of Fig.~\ref{fig04}). For comparison, we plotted again the bottom curve of Fig.~\ref{fig01} on a similar scale as our filtered series in the $2^{nd}$ row of Fig.~\ref{fig04}.  The reconstructed signal is a typical portrait of a beating process as expected for a Be star. Since SSA provides only the most significant eigenvalues found into data, we applied cleanest \citep{fos95} to identify the equivalent frequencies. Cleanest projects the data into an orthogonal subspace of trial functions (sines, cosines, and a constant function), while the projection provided by SSA is onto the data itself.  Figure~\ref{fig04} (bottom) shows the cleanest spectrum of the reconstructed time series.  The cleanest spectrum follow a chi-square distribution with r-1 degrees of freedom, where r is the number of trial functions \citep{fos96}. The frequencies at 2.465 $\rm c\,d^{-1}$ and 2.441 $\rm c\,d^{-1}$ are more than 99.9\% above the significance limit. Figure~\ref{fig05} shows the phase diagrams for the two frequencies found for both smoothed and filtered series.  We had already found those frequencies in the original light curve. To ensure that these frequencies were not introduced by numerical manipulation, we searched for them in each step of our analysis. Figure~\ref{fig06} shows a zoom in the 2.5 $\rm c\,d^{-1}$ region in the cleanest spectrum. From top to bottom we display, the light curve (lc), the lc without orbit perturbations, the lc with interpolated gaps, and the reconstructed time series.  For all the light curves, the signal is above the significance limit. The difference between the two close frequencies is 0.024 $\rm c\,d^{-1}$ above the limit of the frequency resolution, which is 0.018 $\rm c\,d^{-1}$. The amplitude found for both frequencies is 0.14 $\pm$ 0.06 mmag using the algorithm cleanest. The CoRoT photometric performance is 7.7x$10^{-4}$ magnitudes in 1h as faint as V=15.5 \citep{auv03}. \citet{aig09} evaluated the noise per exposure on 2h timescales for the IRa01 and found the photometric performance to be close to the pre-launch specification. They found an empirical relation between 2 h noise and R-magnitude. Applying this empirical relation for the CoRoT-ID 102761769 star and taking all data analyzed (53.47 days), we found that $\sigma_{53.47 days}=$0.01 mmag. However this light curve is particulary noise because of satellite jitter. The standard deviation of the mean from our data is 0.01 mmag. The outliers in the same figure are still caused by the jitter that was not totally eliminated from our analysis. This is why it is difficult to see the 2 frequencies from the $2^{nd}$ row of Fig.~\ref{fig04}.

\begin{figure*}[htp]
\centering
\includegraphics[width=0.6\textwidth,angle=0]{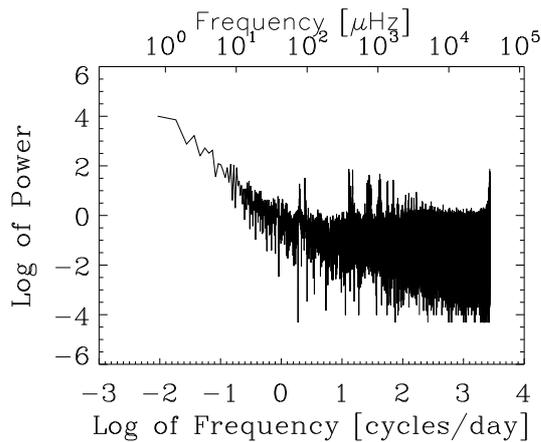}
\caption{Cleanest spectrum of the light curve (Fig. 1 - top). Most of the peaks are harmonics of the satellite orbital period due to the gaps in phase with the orbit (see Fig. 3). Among the significant frequencies, only two of them are not related to the the satellite orbital period harmonics.}
\label{fig02}
\end{figure*}

\begin{figure}[htp]
\centering
\includegraphics[width=0.4\textwidth,angle=0]{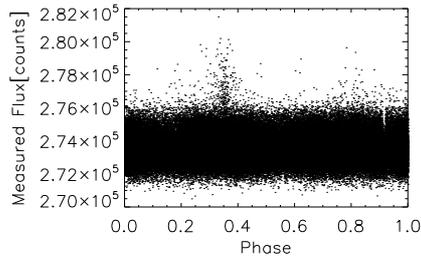}
\caption{Phase diagram of the satellite folded modulo the orbital period (102.9 min). The gaps that appear in phases 0.3 and 0.95 are data eliminated by the automatic pipeline reductions due mainly to satellite jitter. }
\label{fig03}
\end{figure}

\begin{figure*}[htp]
\centering
\includegraphics[width=0.7\textwidth,angle=0]{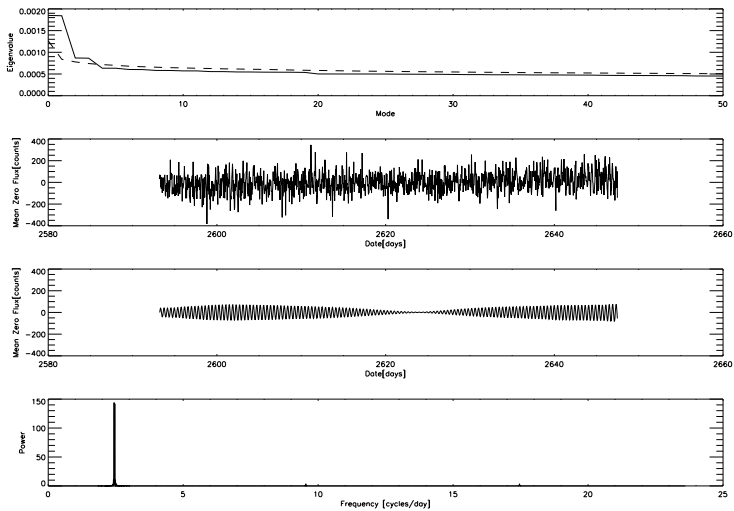}
\caption{Top: The full line shows the singular spectrum and the dashed line shows an average of 100 runs of Monte Carlo simulation. $2^{nd}$ row: Smoothed time series (same as Fig. 1 - bottom). $3^{rd}$ row: Reconstructed time series using only the leading 4 eigenvalues, showing a strong beating effect. Bottom: Fourier transform of the reconstructed series. Note that the frequencies around 2.4 $\rm c\,d^{-1}$ are by far the dominant ones.}
\label{fig04}
\end{figure*}

\begin{figure}[htp]
\centering
\includegraphics[width=3.5 in, height=3.8 in, angle=0]{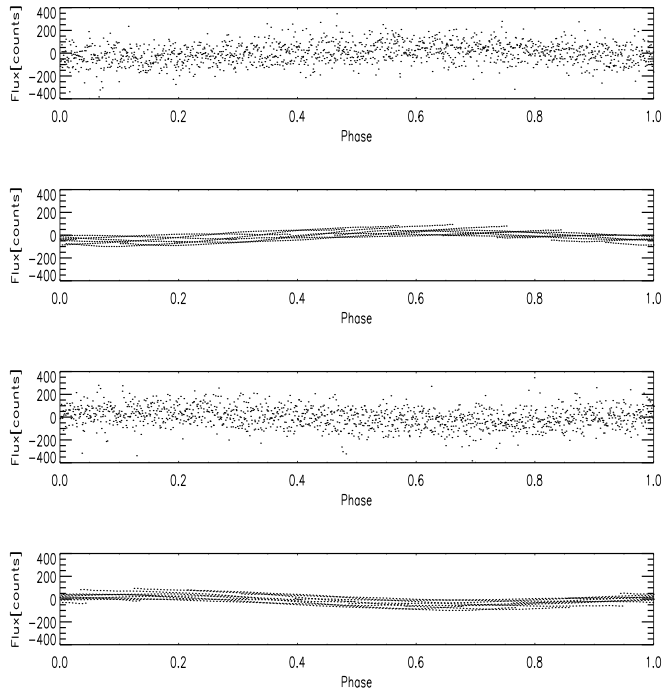}
\caption{Phase diagrams. From top to bottom: Smoothed data with frequency = 2.465 $\rm c\,d^{-1}$, SSA result with frequency = 2.465 $\rm c\,d^{-1}$, smoothed data with frequency = 2.441 $\rm c\,d^{-1}$, SSA result with frequency = 2.441 $\rm c\,d^{-1}$.}
\label{fig05}
\end{figure}

\begin{figure}[htp]
\centering
\includegraphics[width=0.35\textwidth,angle=0]{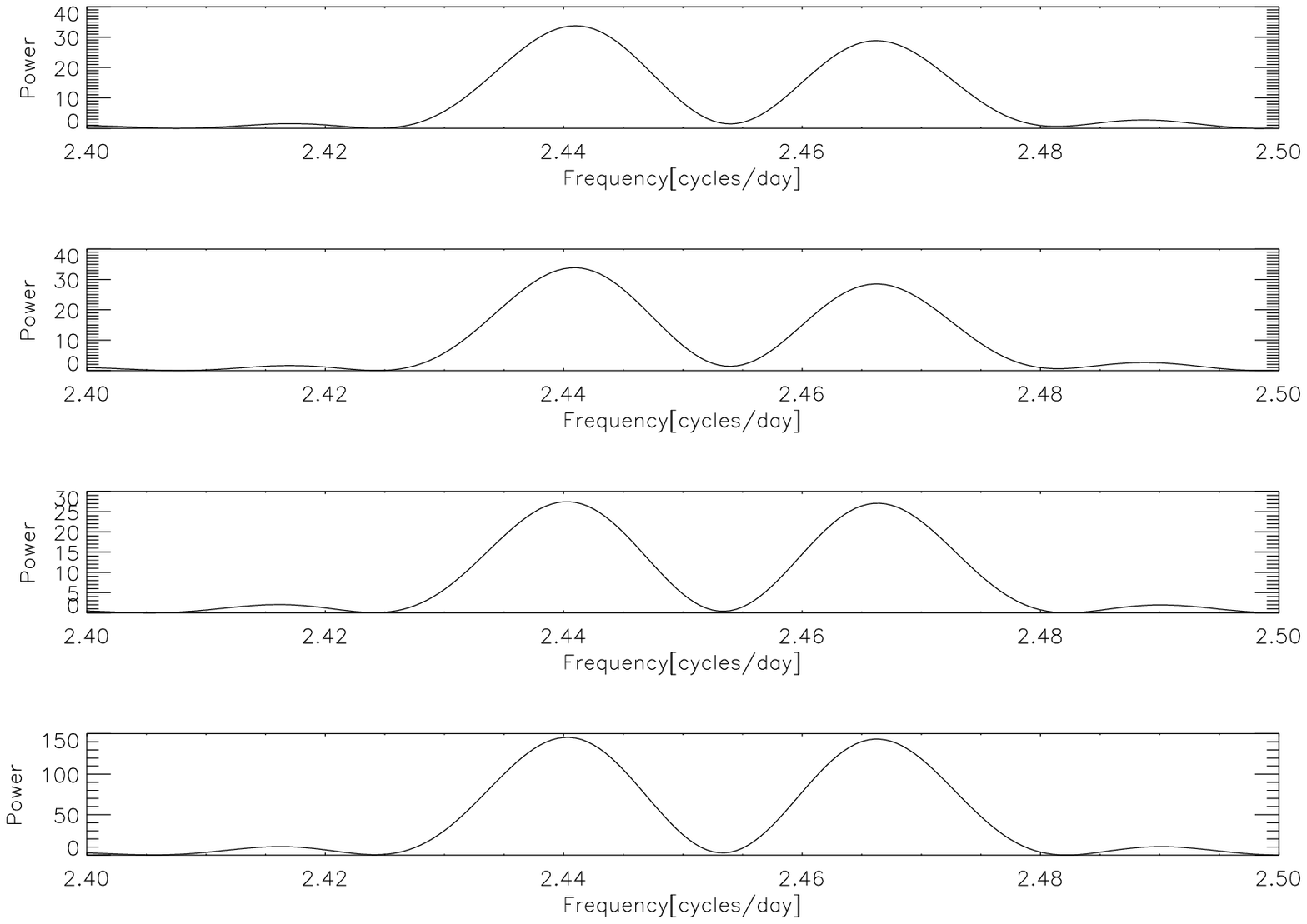}
\caption{Cleanest Spectra between 2.4 and 2.5 $\rm c\,d^{-1}$, from top to bottom: the light curve (lc), lc without orbit perturbations, lc with interpolated gaps, reconstructed time series.}
\label{fig06}
\end{figure}

\section{Spectroscopy ground-based observation}

\begin{figure}[htp]
\centering
\includegraphics[width=0.35\textwidth, angle=0]{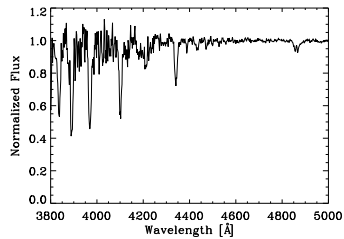}
\caption{Spectroscopic ground-based observation of CoRoT-ID 102761769 star.}
\label{fig07}
\end{figure}

\begin{figure}[htp]
\centering
\includegraphics[width=0.35\textwidth, angle=0]{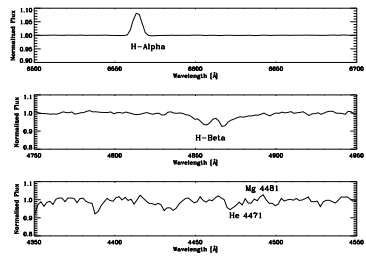}
\caption{Close-ups of interesting spectral regions. From top to bottom: H$\alpha$, H$\beta$, He 4471 $\AA$/Mg 4481 $\AA$}
\label{fig08}
\end{figure}

A spectroscopic observation of the CoRoT-ID 102761769 star was performed at Observatorio del Roque de los Muchachos on March 30 2008 with the Intermediate Dispersion Spectrograph (IDS) at the 2.5m Isaac Newton Telescope.  The spectrum was taken with 300 seconds of exposure time using the 300 l/mm grating centred at 5460 $\AA$ and covering the region from $3337 \AA$ to $8366 \AA$ and resolution of 2 $\AA$ (Fig.~\ref{fig07}). Figure~\ref{fig08} shows views of the H-$\alpha$, H-$\beta$, and He $4471 \AA$/Mg $4481 \AA$ regions.  The reduction was accomplished using IRAF software. The H$\alpha$ line is seen in emission and H$\beta$ is partially filled with emission.  The spectral characteristics are indicative of a spectral type B5-6 IV-Ve based on the EWs of HeI lines, H$\delta$, H$\gamma$, and Mg II ($\lambda 4481 \AA$). The CoRoT catalog indicates for the star an A0V spectral type and luminosity class. This classification was obtained from photometric data using a statistical method \citep{del09}. Our spectral classification based on spectroscopic data provides a more precise classification.
The usual caution must be taken concerning the spectral classification of Be stars, which can be influenced by the circumstellar envelope and the gravitational darkening. The rotation velocity was estimated by evaluating the first zero of the Fourier transform of several He lines. We conclude that the star rotates with V $\sin i = 120 $
$\rm km\,s^{-1}$. This is a maximum value for the projected stellar velocity, because the spectral resolution is about 2 $\AA$, which corresponds to $\approx$ 120 $\rm km\,s^{-1}$.

\section{Discussion}

The two dominant frequencies found in Sect. 3 are $f_1=2.465$ $\rm c\,d^{-1}$ and $f_2=2.441$ $\rm c\,d^{-1}$. Taking $M_*=3.8$\textrm{$M_\odot$} and $R_*=2.6R_\odot$ for a non-flattened B 5.5V star \citep{hug06} and using the expression of \citet{zor05}, the critical rotation velocity is 528 $\rm km\,s^{-1}$ and $\nu_{crit} = 4$ $\rm c\,d^{-1}$. \citet{fre05} found that typical Galactic Be stars rotate on average at 88 \% of their critical rate. If that is the case for CoRoT-ID 102761769, $\nu_{rot} = 3.5 \rm c\,d^{-1}$ and frequencies near $f_1$ and $f_2$ found in this paper are much lower than the probable rotation frequency, which implies that they are true stellar NRP pulsation frequencies. The determined V $\sin i$ and the above critical velocity set the star inclination angle to be $\approx$ 15 degrees. From the H$\alpha$ Ew (6.6 $\AA$) and FWHM (6.2 $\AA$), the relations given by \citet{dac86} allow us to estimate the inner radius of the region emitting the bulk of the line to be $R_\alpha \approx 4.4 R_*$. On the other hand, the absorption cusps seen at the central region of the H$\beta$ line indicate that the inner radius of the H$\beta$ line-forming region is $R_\beta \approx 1.95 R_*$.

\section{Conclusions}

We have analyzed 54.6 days of photometric data of the CoRoT-ID 102761769 target ($V = 13.08$,  B5-6 IV-Ve star) obtained during the IRa01 run. One spectroscopic observation was also performed at Roque de los Muchachos observatory (La Palma, Spain), covering the spectral region $\lambda\lambda$=3337$\AA$-8366$\AA$. The H$\alpha$ line was observed in emission and H$\beta$ was found to be partially filled with emission. The projected rotational velocity is V $\sin i = 120$ $\rm km\,s^{-1}$, as estimated by means of the Fourier transform of HeI line profiles. The time series analysis was based on both cleanest and SSA algorithms. We found two dominant frequencies, $f_1=2.465$ $\rm c\,d^{-1}$ and $f_2=2.441$ $\rm c\,d^{-1}$. A strong beating between these frequencies was uncovered when the data were reconstructed by detecting the four most significant eigenvalues found applying SSA. If the star rotates near the critical frequency, the frequencies $f_1$ and $f_2$ found in this paper are much lower than the probable rotation frequency.

\begin{acknowledgements}
This work is based on observations made with the INT Telescope under the Spanish Instituto de Astrof\'isica de Canarias CAT Service Time. We thank the support astronomers team at the IAC, and, in particular to Rafael Barrena, for making the observations and providing us all the information needed to work on the data. This work was supported by FAPESP (Grant 05/58746-1 and 06/50341-5) and CNPq (Grant 477.813/2007-0 and 480.658/2008-0).
\end{acknowledgements}

\bibliographystyle{aa}

\begin{thebibliography}{}
\bibitem[Aigrain et al. (2009)]{aig09} Aigrain, S., Pont, F., Fressin, F., et al., 2009, A\&A, 506, 425
\bibitem[Allen \& Smith (1996)]{all96} Allen, M.R., \& Smith, L.A. 1996, Journal of Climate, 9, 3373
\bibitem[Auvergne et al. (2003)]{auv03} Auvergne, M., Boisnard, L., Buey, J.-T.M., et al., 2003, in Proc. SPIE 4854, ed. J. C. Blades, \& O. H. W. Siegmund, 170
\bibitem[Auvergne et al. (2009)]{auv09} Auvergne, M., Boisnard, L., Lam-Trong, T., et al., 2009, A\&A, 506, 411
\bibitem[Baade (1982)]{baa82} Baade, D. 1982, A\&A, 105, 65
\bibitem[Baade (1984)]{baa84} Baade, D. 1984, A\&A, 135, 101
\bibitem[Boisnard \& Auvergne (2004)]{boi04} Boisnard, L., \& Auvergne, M. 2004, CoRoT mission engineering, IAC-04-IAFQ. 1.01
\bibitem[Dachs et al. (1986)]{dac86} Dachs, J., Hanuschik, R., Kaiser, D., Rohe, D. 1986, A\&A, 159, 276
\bibitem[Deleuil et al. (2009)]{del09} Deleuil, M., Meunier, J.C., Moutou, C. , et al. 2009, ApJ, 138, 649
\bibitem[Dziembowski et al. (2007)]{dzi07} Dziembowski, W.A., Daszynska-Daszkiewicz L., Pamyatnykh, A.A. 2007, MNRAS 374, 248
\bibitem[Fr\'emat et al. (2005)]{fre05} Fr\'emat, Y., Zorec, J., Hubert, A.-M., Floquet, M. 2005, A\&A, 440,305
\bibitem[Floquet et al. (1992)]{flo92} Floquet, M., Hubert, A.-M., Janot-Pacheco, E., et al 1992, A\&A, 264, 177
\bibitem[Foster (1995)]{fos95} Foster G. 1995, ApJ, 109, 1889
\bibitem[Foster (1996)]{fos96} Foster G. 1996, ApJ, 111, 541
\bibitem[Ghil et al. (2002)]{ghi02}  Ghil, M., Allen, M.R., Dettinge, M.D. 2002, Reviews of Geophysics, 40, 1
\bibitem[Huang \& Gies (2006)]{hug06} Huang, W., Gies, D.R. 2006, ApJ, 648, 451
\bibitem[Kennely et al. (1992)]{ken92} Kennely, E.J., Walker, G.A.H., Merryfield, W.J. 1992, ApJ, 400, L71
\bibitem[Kohoutek \& Wehmeyer (1999)]{koh99}  Kohoutek, L., Wehmeyer, R. 1999, A\&AS, 134, 255
\bibitem[Neiner et al. (2002)]{nei02} Neiner, C., Hubert, A. M., Floquet, M., et al. 2002, A\&A, 388, 899
\bibitem[Porri \& Stalio (1988)]{por88} Porri, A., Stalio, R. 1988, A\&ASS, 75, 371
\bibitem[Porter \& Rivinius (2003)]{por03} Porter, J.M.,\& Rivinius, T. 2003, PASP, 115, 1153
\bibitem[Rivinius et al. (1998)]{riv98} Rivinius, Th., Baade, D., \v{S}tefl, S., et al. 1998, A\&A, 333,125
\bibitem[Rivinius et al. (2001)]{riv01} Rivinius, Th., Baade, D., \v{S}tefl, S., Maintz, M. 2001, A\&A, 369, 1058
\bibitem[Smith (1977)]{smi77} Smith, M.A. 1977, ApJ, 215, 574
\bibitem[Robertson \& Jordan (1989)]{rob89} Robertson, T.H., Jordan, T.M. 1989, AJ, 98, 1354
\bibitem[Steele et al. (1999)]{ste99} Steele, I.A., Negueruela, I., Clarck, J.S. 1999, A\&AS 137, 147
\bibitem[Vogt \& Penrod (1983)]{vog83} Vogt, S.S., \&Penrod G.D. 1983, ApJ, 275,661
\bibitem[Walker et al. (2005)]{wal05} Walker, G.A.H., Kuschnig, R., Mathews, J.M., et al. 2005, ApJ, 635, L77
\bibitem[Zorec et al. (2005)]{zor05} Zorec, J., Fr\'emat, Y., Cidale, L. 2005, A\&A, 441, 235

\end{thebibliography}

\end{document}